\documentclass[aps,nofootinbib,prd,preprint,tightenlines,amsmath,amssymb]{revtex4-2}

\usepackage{bm}
\usepackage[colorlinks,anchorcolor=black,citecolor=violet,linkcolor=blue]{hyperref}
\usepackage{graphicx}
\usepackage{multirow,slashed}
\usepackage{xcolor}

\begin{document}
\baselineskip=17pt \parskip=5pt

\preprint{NCTS-PH/2006}

\title{Seeking massless dark photons in the decays of charmed hadrons}

\author{Jhih-Ying Su$^1$ and Jusak Tandean$^{1,2}$ \smallskip \\ \it
$^1$Department of Physics, National Taiwan University, Taipei 106, Taiwan \\
$^2$Physics Division, National Center for Theoretical Sciences, Hsinchu 300, Taiwan \bigskip \\
\large\rm Abstract \medskip \\
\begin{minipage}{\textwidth} \baselineskip=17pt \parindent=3ex \small
A massless dark photon could affect standard-model particles only via higher-dimensional
operators and would therefore have eluded recent searches for its massive counterpart,
which were based on the assumption that the latter had renormalizable interactions with
known fermions due to gauge kinetic mixing.
In this study we entertain the possibility that the massless dark photon has nonnegligible
flavor-changing dipole-type couplings with the $u$ and $c$ quarks, giving rise to the decays
of charmed hadrons into a lighter hadron plus missing energy carried away by the dark photon.
We propose to investigate decays of this kind, especially those in which the parents
are the charmed pseudoscalar-mesons $D^+$,~$D^0$, and $D_s^+$ and singly charmed baryons
$\Lambda_c^+$, $\Xi_c^+$, and $\Xi_c^0$.
Employing a simplified new-physics model satisfying the relevant constraints, we find that
the branching fractions of these processes could be as large as several times $10^{-5}$.
This suggests that one or more of them might in the near future fall within reach of
the ongoing Belle II and BESIII experiments.
Since the same underlying operators are responsible for all of these transitions, detecting
one of them automatically implies particular predictions for the others, allowing for
additional experimental checks on the massless-dark-photon scenario.
\end{minipage}}


\maketitle

\section{Introduction\label{intro}}

Attempts to address longstanding open questions in physics, such as the nature and origin of
neutrino mass and the particle identity of cosmic dark matter, have increasingly postulated
the existence of a dark sector beyond the standard model (SM).
It is reasonable to expect that the new dark sector not only provides resolutions to some of
those major puzzles but also furnishes extra ingredients which will facilitate further
empirical access to it.
Among the most attractive ones is a~dark Abelian gauge group, U(1)$_D$, under which all SM
fields are singlets.
This symmetry may be spontaneously broken or stay unbroken, causing the associated gauge boson,
the dark photon, to gain mass or remain massless, respectively.

Whether it is massive or massless, the hope is that the dark photon can somehow communicate
with the SM as well as connect it to other constituents of the dark side,
leading to interesting and potentially observable consequences.
These possibilities have received a good deal of theoretical attention in the past few
decades~\cite{Okun:1982xi,Georgi:1983sy,Holdom:1985ag,Foot:1991kb,delAguila:1995rb,
Dobrescu:2004wz,Gabrielli:2016cut,Hoffmann:1987et,Fargion:2005ep,Fabbrichesi:2017vma,
Su:2020xwt,Su:2019ipw,Ackerman:2009mha,Barger:2011mt,Chiang:2016cyf,He:2017zzr,Zhang:2018fbm,
Pospelov:2008zw,Jaeckel:2010ni,Essig:2013lka,Alexander:2016aln,Fabbrichesi:2020wbt} and
stimulated numerous dedicated hunts for dark photons~\cite{Essig:2013lka,Alexander:2016aln,
Fabbrichesi:2020wbt,Batley:2015lha,Ablikim:2017aab,Aaij:2017rft,Anastasi:2018azp,Ablikim:2018bhf,
CortinaGil:2019nuo,NA64:2019imj,Tanabashi:2018oca}, albeit still with negative outcomes to date.
Most of these efforts have focused on the massive dark photon, $A'$, which can couple directly
to SM fermions via the renormalizable operator $\epsilon eA_\mu'J_{\textsc{em}}^\mu$ involving
the electromagnetic current $eJ_{\textsc{em}}$ and a small constant $\epsilon$ brought about by
the kinetic mixing between the dark and SM Abelian gauge fields~\cite{Pospelov:2008zw,
Jaeckel:2010ni,Essig:2013lka,Alexander:2016aln,Fabbrichesi:2020wbt}.
In the presence of this coupling, $A'$ could be produced in the decays or scatterings of
ordinary leptons and quarks, including those of hadrons, and it could decay into
electrically charged fermions or mesons.
Owing to these properties of $A'$, the measurements looking for it have been able to acquire
limits on $\epsilon$ over various ranges of its mass~\cite{Jaeckel:2010ni,Essig:2013lka,
Alexander:2016aln,Fabbrichesi:2020wbt,Batley:2015lha,Ablikim:2017aab,Aaij:2017rft,
Anastasi:2018azp,Ablikim:2018bhf,CortinaGil:2019nuo,NA64:2019imj,Tanabashi:2018oca}.

If the dark photon is massless, the situation is very different but no less interesting.
In this case, one can always define a linear combination of the dark and SM U(1) gauge fields
that has no renormalizable interactions with the SM and is identified with the massless
dark photon~\cite{Holdom:1985ag,Dobrescu:2004wz}, which we denote by $\bar\gamma$.
The implications are that it has no direct couplings to SM fermions, in contrast to
its massive counterpart, and that therefore restrictions inferred from the aforesaid quests for
$A'$ do not apply to $\bar\gamma$.
However, the latter can still affect the SM sector through higher-dimensional operators
induced by loop diagrams containing particles charged under U(1)$_D$ and also coupled to SM
fields \cite{Dobrescu:2004wz,Gabrielli:2016cut,Hoffmann:1987et}.
This means that there may be more efficient ways to probe $\bar\gamma$, which are worth
pursuing and several of which will be explored in this work.

In the absence of other particles beyond the SM lighter than the electroweak scale,
the effective interactions of the massless dark photon with SM members can be described
by operators which respect the SM gauge group and the unbroken U(1)$_D$.
At leading order the couplings of $\bar\gamma$ to quarks are of dipole type and given by
the gauge-invariant Lagrangian \cite{Dobrescu:2004wz}
\begin{align} \label{Lnp0}
{\cal L}_{\textsc{np}}^{} & \,=\, \frac{1}{\Lambda_{\textsc{np}}^2} \Big(
{\cal C}_{jk}^{}\, \overline{q_j^{}} \sigma^{\mu\nu} d_k^{} H
+ {\cal C}_{jk}'\, \overline{q_j^{}} \sigma^{\mu\nu} u_k^{} \tilde H
\,+\, {\rm H.c.} \Big) \bar F_{\mu\nu}^{} \,, ~~~ ~~~~
\end{align}
where $\Lambda_{\textsc{np}}$ denotes an effective heavy mass, the ${\cal C}$s are dimensionless
coefficients which are generally complex, $q_j^{}$ and $d_k^{}$ ($u_k$) represent a left-handed
quark doublet and right-handed down(up)-type quark singlet, respectively, under
the SU(2)$_L$ gauge group, $H$ stands for the SM Higgs doublet,
\,$\tilde H=i\tau_2^{}H^*$\, with $\tau_2^{}$ being the second Pauli matrix,
\,$\bar F_{\mu\nu}=\partial_\mu\bar A_\nu-\partial_\nu\bar A_\mu$\, is the dark photon's
field-strength tensor, \,$\sigma^{\mu\nu}=i[\gamma^\mu,\gamma^\nu]/2$,\, and
summation over family indices \,$j,k=1,2,3$\, is implicit.
Both $\Lambda_{\textsc{np}}$ and the ${\cal C}$s depend on the details of the underlying
new physics (NP), and in general ${\cal C}_{jk}^{}$ and ${\cal C}_{jk}'$ are not necessarily
related to one another.

Here we concern ourselves with flavor-changing neutral current (FCNC) transitions which arise
from the operators in Eq.\,(\ref{Lnp0}), with quarks of the first two generations,
and contribute to the decays of hadrons with missing energy.
In the SM the corresponding reactions have an unobserved neutrino pair in the final state and
are greatly suppressed, as they proceed from loop diagrams and are subject to
the Glashow-Iliopoulos-Maiani mechanism~\cite{Buchalla:1995vs,Burdman:2001tf}.
It follows that these FCNC hadron decays may be promising places to seek signs of NP.
If ${\cal L}_{\textsc{np}}$ could impact these processes significantly, a number of them might
have rates that are amplified far above their SM expectations to values within the sensitivity
ranges of ongoing high-intensity flavor experiments like NA62~\cite{NA62:2017rwk},
KOTO~\cite{Ahn:2018mvc}, BESIII~\cite{Li:2016tlt,Ablikim:2019hff,Asner:2008nq}, and
Belle II~\cite{Kou:2018nap}.

The scenario in which the massless dark photon, $\bar\gamma$, possesses flavor-violating
interactions with the $d$ and $s$ quarks, via the ${\cal C}_{jk}$ portion of
${\cal L}_{\textsc{np}}$, has already been investigated previously.
They translate into the FCNC transitions of strange hadrons with missing
energy carried away by $\bar\gamma$.
As pointed out in Refs.\,\cite{Fabbrichesi:2017vma,Su:2020xwt} (\cite{Su:2019ipw}), decays of
this kind in the kaon (hyperon) sector have branching fractions that are permitted by current
constraints to increase to levels which might be discoverable soon by NA62 and KOTO (BESIII).

In the present paper we concentrate on the possibility that the massless dark photon has
flavor-changing interactions with the $u$ and $c$ quarks, which originate from
the ${\cal C}_{jk}'$ parts in~Eq.\,(\ref{Lnp0}).
After electroweak symmetry breaking, in the mass basis of the up-type quarks we can express
the relevant terms as
\begin{align} \label{Lucg}
{\cal L}_{uc\bar\gamma}^{} & \,=\, \overline u \big( {\mathbb C}
+ \gamma_5^{} {\mathbb C}_5^{} \big) \sigma^{\mu\nu} c\, \bar F_{\mu\nu}^{}
\,+\, {\rm H.c.} \,, ~~~ ~~~~
\end{align}
and so
\,$\mbox{\small$\mathbb C$} =
\Lambda_{\textsc{np}}^{-2}\big({\cal C}_{12}'+{\cal C}_{21}^{\prime*}\big)v/\sqrt8$\,
and
\,$\mbox{\small$\mathbb C$}_5^{} =
\Lambda_{\textsc{np}}^{-2}\big({\cal C}_{12}'-{\cal C}_{21}^{\prime*}\big)v/\sqrt8$\,
are parameters which have the dimension of inverse mass and are determined by the specifics
of the ultraviolet-complete model, with $v\simeq246$\,\,GeV being the Higgs vacuum expectation
value.
In the next few sections, we analyze the implications for the FCNC decays of several charmed
hadrons into two-body final states each consisting of a lighter hadron and $\bar\gamma$.
The corresponding SM contributions are the three-body modes with a neutrino pair which have
tiny branching fractions~\cite{Burdman:2001tf} and hence virtually negligible.
On the other hand, the two-body decays with the massless dark photon due to
${\cal L}_{uc\bar\gamma}$ could have relatively substantial rates, but there have been no
direct searches for them yet, as far as we can tell~\cite{Tanabashi:2018oca}.
Therefore, it is hoped that our study will motivate empirical efforts to pursue them,
which may shed light on the existence of $\bar\gamma$ or at least come up with bounds
on {\small$\mathbb C$} and {\small$\mathbb C$}$_5$.

Since a clean environment and sizable luminosity are essential for such endeavors,
it is timely that we now have running flavor factories which are potentially
well-suited for them, namely BESIII and Belle II.
In these experiments, two charmed hadrons can be created in flavor-conjugate states from
each \,$e^+e^-\to c\bar c$\, event above the pertinent threshold.
Moreover, one of the charmed hadrons can be fully reconstructed, and subsequently energy-momentum
conservation can be invoked to look for the decay of the other involving missing
energy~\cite{Kou:2018nap,Asner:2008nq,Lai:2016uvj}.
In the future, this procedure would be implemented, likely with much improved sensitivity reach, at
the proposed super charm-tau factories~\cite{Luo:2018njj,Barnyakov:2020vob} and Circular
Electron Positron Collider (CEPC) operated as a $Z$-boson factory~\cite{CEPCStudyGroup:2018ghi}.

The rest of the paper is organized in the following way.
In Sec.\,\,\ref{Dmesons} we deal with the decays of the lightest charmed hadrons, namely
the pseudoscalar mesons $D^+$, $D^0$, and $D_s^+$, into a charmless meson plus $\overline\gamma$.
For good measure, we include \,$D^0\to\gamma\bar\gamma$,\, which is also induced by
${\cal L}_{uc\bar\gamma}$ and emits an ordinary photon, $\gamma$, instead of a meson.
In Sec.\,\,\ref{Bc} we address the decays of singly charmed baryons $\Lambda_c^+$, $\Xi_c^+$, and
$\Xi_c^0$ into a charmless baryon and $\overline\gamma$.
Testing these meson and baryon processes would be most feasible at $e^+e^-$ machines,
as described in the last paragraph.
Charmed hadrons containing more than one heavy quark could undergo analogous reactions, but are
either much less likely or too heavy to be produced at existing or upcoming $e^+e^-$ facilities,
and so these transitions probably will not be probed for some time.
Nevertheless, since the required information is available, for completeness we evaluate in
Secs.\,\,\ref{Xcc} and \ref{Bcmeson} the decays of doubly charmed baryon $\Xi_{cc}^{++}$ and
bottom, charmed pseudoscalar-meson $B_c^+$, respectively.
Finally, in Sec.\,\ref{concl} we give our conclusions.
The Appendix supplies extra details on the baryonic matrix elements of the dipole operators
in~${\cal L}_{uc\bar\gamma}$.

\section{Decays of charmed mesons\label{Dmesons}}

The interactions in ${\cal L}_{uc\bar\gamma}$ give rise to the FCNC decays of the lightest
charmed mesons into a~charmless meson and the dark photon.
Given that the decay of a spinless particle into another spinless particle plus a massless gauge
boson is forbidden by angular-momentum conservation and gauge invariance, in most of this section
we consider modes where the initial particle is the charmed pseudoscalar-meson $D^+$ or~$D^0$ or
the charmed, strange pseudoscalar-meson $D_s^+$, whereas the daughter particles consist of
a charmless vector-meson and $\bar\gamma$.
In particular, we examine \,$D^+\to\rho^+\bar\gamma$,\, $D^0\to\rho^0\bar\gamma$,\,
$D^0\to\omega\bar\gamma$,\, and \,$D_s^+\to K^{*+}\bar\gamma$.\,
As for \,$D^0\to\gamma\bar\gamma$,\, we will treat it toward the end of the section.

For \,$D^+\to\rho^+\bar\gamma$\, the amplitude ${\cal M}_{D^+\to\rho^+\bar\gamma}$ contains
the mesonic matrix elements
$\langle\rho^+|\overline u\sigma^{\mu\nu}c|D^+\rangle$ and
$\langle\rho^+|\overline u\sigma^{\mu\nu}\gamma_5^{}c|D^+\rangle$, the general formulas for
which are long known in the literature~\cite{Isgur:1990kf,Wu:2006rd,Melikhov:2000yu}.
In ${\cal M}_{D^+\to\rho^+\bar\gamma}$ these matrix elements are to be contracted with
the outgoing dark photon's momentum $\bar q$ and polarization vector $\bar\varepsilon$.
These 2 four-vectors fulfill the gauge requirement \,$\bar\varepsilon\cdot\bar q=0$\, and
the masslessness condition \,$\bar q^2=0$\, owing to $\bar\gamma$ being on shell.
Accordingly, we can write
\begin{align} \label{<r|usc|D>}
\langle\rho^+(k)|\overline u\sigma^{\mu\nu}c|D^+(k+\bar q)\rangle\,
\bar\varepsilon_\mu^*\bar q_\nu^{}
& \,=\, 2i f_{D^+\rho^+}^{}\, \epsilon^{\eta\tau\mu\nu} \varepsilon_\eta^* k_\tau^{}
\bar\varepsilon_\mu^* \bar q_\nu^{} \,,
\nonumber \\
\langle\rho^+(k)|\overline u\sigma^{\mu\nu}\gamma_5^{}c|D^+(k+\bar q)\rangle\,
\bar\varepsilon_\mu^* \bar q_\nu^{}
& \,=\, 2 f_{D^+\rho^+}^{} \big( \varepsilon^*\!\cdot\!\bar q\; \bar\varepsilon^*\!\cdot\!k
- \varepsilon^*\!\cdot\!\bar\varepsilon^*\, k\!\cdot\!\bar q \big) \,,
\end{align}
where $k$ and $\varepsilon$ stand for the $\rho^+$ momentum and polarization vector,
respectively, $k+\bar q$ is the $D^+$ momentum, and the same constant $f_{D^+\rho^+}$
parametrizes form-factor effects at \,$\bar q^2=0$\, in the two equations,
which are related~\cite{Isgur:1990kf} by virtue of the identity
\,$2i\sigma^{\eta\kappa}\gamma_5^{}=\epsilon^{\eta\kappa\nu\tau}\sigma_{\nu\tau}$.\,
They lead to
\begin{align} \label{MD2rg}
{\cal M}_{D^+\to\rho^+\bar\gamma}^{} & \,=\, 4 f_{D^+\rho^+}^{} \Big[ \epsilon^{\eta\kappa\nu\tau}
\varepsilon_\eta^* \bar\varepsilon_\kappa^* k_\nu^{} \bar q_\tau^{}\, {\mathbb C}
+ i \big( \varepsilon^*\!\cdot\!\bar q\; \bar\varepsilon^*\!\cdot\!k
- \varepsilon^*\!\cdot\!\bar\varepsilon^*\,k\!\cdot\!\bar q \big) {\mathbb C}_5^{} \Big] \,, ~~~~~~~
\end{align}
which is U(1)$_D$-gauge invariant and from which we obtain the branching fraction
\begin{align} \label{BD2rg}
{\cal B}\big(D^+\to\rho^+\bar\gamma\big) & \,=\, \frac{\tau_{D^+}^{}\, f_{D^+\rho^+}^2
\big(m_{D^+}^2-m_{\rho^+}^2\big)\raisebox{1pt}{$^3$}}{2 \pi m_{D^+}^3}
\big( |{\mathbb C}|^2 + |{\mathbb C}_5|^2 \big) \,, ~~~ ~~~~
\end{align}
where $\tau_{D^+}$ represents the lifetime of $D^+$ and $m_{\cal X}^{}$ denotes the mass of
$\cal X$.
The corresponding quantities for \,$D^0\to\rho^0\bar\gamma,\omega\bar\gamma$\, and
\,$D_s^+\to K^{*+}\bar\gamma$\, have analogous expressions.

For numerical calculations, we employ
\begin{align} \label{gDV}
f_{D^+\rho^+}^{} & \,=\, \sqrt2\, f_{D^0\rho^0} \,=\, 0.658^{+0.038}_{-0.031} \,, &
f_{D^0\omega}^{} & \,=\, 0.610^{+0.036}_{-0.030} \,, &
f_{D_s^+K^{*+}}^{}   & \,=\, 0.639^{+0.042}_{-0.044} \,, ~~~
\end{align}
which have been estimated in Ref.\,\cite{Wu:2006rd} with light-cone sum rules in
the framework of heavy-quark effective field theory.
The relation between $f_{D^+\rho^+}$ and $f_{D^0\rho^0}$ in Eq.\,(\ref{gDV}) follows from
the quark flavor contents \,$\rho^+\sim u\bar d$\, and \,$\rho^0\sim(u\bar u-d\bar d)/\sqrt2$.\,
For comparison, an earlier analysis within a constituent quark model~\cite{Melikhov:2000yu}
yielded \,$f_{D^+\rho^+}=0.66$\, and \,$f_{D_s^+K^{*+}}=0.71$\,
with uncertainties of around 10\%, which are compatible with their newer counterparts
in Eq.\,(\ref{gDV}).
Additional input parameters are the empirical $D^{+,0}$ and $D_s^+$ lifetimes and masses and 
the light meson masses, namely \,$m_{\rho^+}=775.11(34)$,\, $m_{\rho^0}=775.26(25)$,\,
$m_\omega=782.65(12)$,\, and \,$m_{K^{*+}}=895.5(8)$,\, in units of MeV,
all from Ref.\,\cite{Tanabashi:2018oca}.
Thus, the only remaining unknowns are the coefficients $\mathbb C$ and ${\mathbb C}_5$,
which depend on the details of the NP model.

These processes can serve as valuable probes of the massless dark photon if their
rates are not highly suppressed.
Clearly, this can happen provided that one or both of $\mathbb C$ and ${\mathbb C}_5$ are not
too small in size.
At the moment, there are no model-independent restrictions on them, as there are still
no data on these charmed-hadron decays~\cite{Tanabashi:2018oca}.
It turns out that there is at least one NP model in the recent literature~\cite{Gabrielli:2016cut}
which offers some nonnegligible viable values of these coefficients.
Those numbers can then serve as benchmarks to illustrate how these charmed-hadron decays may
shed light on the dark photon's existence.

In the NP scenario of Ref.\,\cite{Gabrielli:2016cut} the $c\to u\bar\gamma$ operators arise
from loop diagrams involving massive fermions which are singlets under the SM gauge group and
heavy new scalar bosons which carry some of the SM gauge charges.
The new fermions and bosons are all charged under U(1)$_D$ and have Yukawa-like interactions
with the $u$ and $c$ quarks.
This allows for the construction of the dimension-five operators in Eq.\,(\ref{Lucg}).
Since in this paper we are mainly interested in the implications of these effective
interactions for the charmed-hadron decays, we will not dwell further on the specifics of
the underlying NP model.
Rather, we will simply adopt the relevant results available from Ref.\,\cite{Gabrielli:2016cut}
at face value and apply them to the determination of the rates of the charmed-hadron processes.

Particularly, as elaborated in Ref.\,\cite{Gabrielli:2016cut}, the branching fraction of
the inclusive transition \,$c\to u\bar\gamma$\, could be as much as about $10^{-4}$,
with the pertinent constraints, from dark matter and vacuum stability considerations,
having been taken into account.
Accordingly, for the purposes of our study we can take
\,${\cal B}(c\to u\bar\gamma)<5\times10^{-5}$,\, which translates into\footnote{This has been
found by using
\,${\cal B}(c\to u\bar\gamma) = 12\pi^2 {\cal B}(c\to\ell^+X)_{\rm exp} \big( 1/|\Lambda_L|^2
+ 1/|\Lambda_R|^2 \big)/\big[ G_{\rm F}^2m_c^2|V_{\rm cs}|^2 \textsl{\texttt f}_1(m_s/m_c) \big]$
upon adding the factor $\pi^2$ missing from Eq.\,(88) in \cite{Gabrielli:2016cut}, with
\,${\cal B}(c\to\ell^+X)_{\rm exp}=0.096$\, from the $D^+$ listing in~\cite{Tanabashi:2018oca},
$G_{\rm F}$ being the Fermi constant, \,$m_c^{}=1.67$\,\,GeV,
\,$|V_{cs}|=0.986$\, \cite{Gabrielli:2016cut},
\,$\textsl{\texttt f}_1(x)=1-8x^2+8x^6-x^8-24x^4\ln x$,\, and
\,$m_s/m_c\simeq0.0853$ \cite{Tanabashi:2018oca}, along with the relation
\,$1/|\Lambda_L|^2+1/|\Lambda_R|^2=8|{\mathbb C}|^2 + 8|{\mathbb C}_5|^2$.\medskip}
\begin{align} \label{C2max}
|{\mathbb C}|^2 + |{\mathbb C}_5|^2 & \,<\, \frac{1.9\times10^{-16}}{\rm GeV^2} \,.
\end{align}
It might seem that this is very optimistic because ${\cal B}(c\to u\bar\gamma)$ in
Ref.\,\cite{Gabrielli:2016cut} ranges from $10^{-13}$ to $10^{-4}$, as Table\,VI and Fig.\,6
therein show.
However, given that direct-search bounds on the couplings are currently absent, it is possible
that there are other NP models which can generate numbers larger than that in Eq.\,(\ref{C2max}).
Furthermore, model-independently,
\,$|{\mathbb C}|^2+|{\mathbb C}_5|^2\sim10^{-15}\rm/GeV^2$\, would even still be permitted at
present, being compatible with the caps on the branching fractions of yet-unobserved $D^{+,0}$
and $D_s^+$ decays.\footnote{With \,$|{\mathbb C}|^2+|{\mathbb C}_5|^2=10^{-15}\rm/GeV^2$,\,
the branching-fraction maxima in Eq.\,(\ref{D2Vg'limits}) would be {\footnotesize$\sim$\,}5
times greater.
The increased results are well under the caps on the branching fractions of yet-unobserved
decay channels of $D^{+,0}$ and $D_s^+$, which exceed 1\%, as can be inferred from the data on
their observed ones~\cite{Tanabashi:2018oca}.}
Therefore, what we have in Eq.\,(\ref{C2max}) is a reasonable reference value with which
to make benchmark predictions that are potentially testable by searches in the near future.

The rates of all the decays which will be discussed shortly are proportional to
$|{\mathbb C}|^2 + |{\mathbb C}_5|^2$.
Consequently, if hunts for these transitions come up with limits stronger than the branching
fractions we compute below based on Eq.\,(\ref{C2max}), the implied restraints are germane solely
to this combination of the coefficients.
To investigate $\mathbb C$ and ${\mathbb C}_5$ separately would require measuring more complicated
observables, such as the angular distributions of the decay products, but this step will be
worth taking only after one or more of these decays are discovered.

Incorporating the central values of the aforementioned input parameters and Eq.\,(\ref{C2max})
into the meson branching fractions, we arrive at
\begin{align} \label{D2Vg'limits}
{\cal B}\big(D^+\to\rho^+\bar\gamma\big) & \,=\, 4.04\times10^{11}\, \tilde{\textsc c}{}^2
\,<\, 7.7\times10^{-5} \,,
\nonumber \\
{\cal B}(D^0\to\rho^0\bar\gamma) & \,=\, 7.88\times10^{10}\, \tilde{\textsc c}{}^2
\,<\, 1.5\times10^{-5} \,,
\nonumber \\
{\cal B}(D^0\to\omega\bar\gamma) & \,=\, 1.34\times10^{11}\, \tilde{\textsc c}{}^2
\,<\, 2.5\times10^{-5} \,,
\nonumber \\
{\cal B}(D_s^+\to K^{*+}\bar\gamma) & \,=\, 1.89\times10^{11}\, \tilde{\textsc c}{}^2
\,<\, 3.6\times10^{-5} \,, ~~~ ~~~~
\end{align}
where, for conciseness, we have defined
\begin{align} \label{c^2}
\tilde{\textsc c}{}^2 & \,=\, \big(|{\mathbb C}|^2+|{\mathbb C}_5|^2\big) \rm\,GeV^2 \,. 
\end{align}
Interestingly, these numbers are comparable to the latest limit
\,${\cal B}(D^0\to\mbox{invisibles})<9.4\times10^{-5}$ at 90\% confidence
level \cite{Tanabashi:2018oca} set by the Belle Collaboration \cite{Lai:2016uvj}.
This suggests that one or more of the results in Eq.\,(\ref{D2Vg'limits}) may be within reach
of the ongoing BESIII \cite{Ablikim:2019hff} and Belle II \cite{Kou:2018nap}
experiments.\footnote{We may take the Belle finding on
${\cal B}(D^0\to\mbox{invisibles})$ as a guide for making a rough guess on whether
Eq.\,(\ref{D2Vg'limits}) may be accessible to Belle II.
The detector performance of Belle II is similar to that of Belle, but the accumulated
integrated luminosity will be 50 times bigger~\cite{Kou:2018nap}.
Thus, if the reconstruction efficiency of the $\rho$ meson is 20\% or higher at Belle II,
we expect it to have the sensitivity
\,${\cal B}(D\to\rho\bar\gamma)\sim9.4\times10^{-5}/(0.2\times50)\sim10^{-5}$\, or better,
indicating that it may be able to test the results in Eq.\,(\ref{D2Vg'limits}).}
If not, significantly improved probes would probably be available from proposed facilities like
the super charm-tau factories~\cite{Luo:2018njj,Barnyakov:2020vob}
and CEPC \cite{CEPCStudyGroup:2018ghi}.

For \,$D^0\to\gamma\bar\gamma$,\, we need the matrix elements
\begin{align}
\langle\gamma(k)|\overline u\sigma^{\mu\nu}c|D^0(k+\bar q)\rangle\,
\bar\varepsilon_\mu^*\bar q_\nu^{}
& \,=\, i e f_{D^0\gamma}^{}\, \epsilon^{\eta\tau\mu\nu} \check\varepsilon_\eta^* k_\tau^{}
\bar\varepsilon_\mu^* \bar q_\nu^{} \,,
\nonumber \\
\langle\gamma(k)|\overline u\sigma^{\mu\nu}\gamma_5^{}c|D^0(k+\bar q)\rangle\,
\bar\varepsilon_\mu^* \bar q_\nu^{}
& \,=\, e f_{D^0\gamma}^{}
\big( \check\varepsilon^*\!\cdot\!\bar q\; \bar\varepsilon^*\!\cdot\!k
- \check\varepsilon^*\!\cdot\!\bar\varepsilon^*\, k\!\cdot\!\bar q \big) \,, ~~~
\end{align}
which have also been considered in the literature \cite{Geng:2000if,Hazard:2017udp} and involve
the electric charge $e$, the photon's polarization vector $\check\varepsilon$,
and the form-factor parameter $f_{D^0\gamma}$.
Therefore, the decay amplitude satisfies both the U(1)$_D$ and electromagnetic gauge symmetries.
The corresponding branching fraction is
\begin{align} \label{BD2gg'}
{\cal B}\big(D^0\to\gamma\bar\gamma\big) & \,=\, \frac{\alpha_{\rm e}^{}}{2}\, \tau_{D^0}^{}
f_{D^0\gamma}^2 m_{D^0}^3 \big( |{\mathbb C}|^2 + |{\mathbb C}_5|^2 \big) \,, ~~~
\end{align}
where \,$\alpha_{\rm e}=e^2/(4\pi)=1/137$.\,
Hence including \,$f_{D^0\gamma}=0.33$\, from Ref.\,\cite{Hazard:2017udp} and
Eq.\,(\ref{C2max}) leads to
\begin{align} \label{D2gg'limits} &
{\cal B}(D^0\to\gamma\bar\gamma) \,=\, 1.61\times10^9\, \tilde{\textsc c}{}^2
\,<\, 3.1\times10^{-7} \,. ~~~~~
\end{align}
This is roughly 2 orders of magnitude below its counterparts in Eq.\,(\ref{D2Vg'limits}),
which is due to the presence of $\alpha_{\rm e}$.

\section{Decays of singly charmed baryons\label{Bc}}

In the baryon sector, the \,$c\to u\bar\gamma$\, transition in Eq.\,(\ref{Lucg}) also causes
charmed baryons to undergo \,$\Delta C=1$\, decays into a lighter baryon plus the dark photon.
In this section, we focus on the reactions of the singly charmed baryons $\Lambda_c^+$, $\Xi_c^+$,
and $\Xi_c^0$,\, which have spin parity \,$J^P=1/2^+$,  make up a flavor SU(3) antitriplet,
and decay weakly \cite{Tanabashi:2018oca}.
We examine in particular the modes \,$\Lambda_c^+\to p\bar\gamma$\, and
\,$\Xi_c^{+,0}\to\Sigma^{+,0}\bar\gamma$,\, which BESIII and Belle II could also pursue.

To derive the amplitude ${\cal M}_{\Lambda_c^+\to p\bar\gamma}$ for \,$\Lambda_c^+\to p\bar\gamma$\,
from the short-distance interaction described by ${\cal L}_{uc\bar\gamma}$, we need to know
the matrix elements \,$\langle p|\overline u\sigma^{\mu\nu}c|\Lambda_c^+\rangle$\, and
\,$\langle p|\overline u\sigma^{\mu\nu}\gamma_5^{}c|\Lambda_c^+\rangle$.\,
Contracting them with the dark photon's momentum $\bar q$ and polarization vector
$\bar\varepsilon$, we can write
\begin{align} \label{<p|usc|Lc>}
\langle p|\overline u\sigma^{\mu\nu} \big(1,\gamma_5^{}\big) c|\Lambda_c^+\rangle\,
\bar\varepsilon_\mu^* \bar q_\nu^{}
& \,=\, f_{\Lambda_c^+p}^{}\, \overline{\textsl{\texttt U}_p^{}}\, \sigma^{\mu\nu}
\big(1,\gamma_5^{}\big) \textsl{\texttt U}_{\Lambda_c}^{} \bar\varepsilon_\mu^*\bar q_\nu^{} \,, ~~~
\end{align}
where $\textsl{\texttt U}_{\Lambda_c}$ and $\textsl{\texttt U}_p$ stand for the Dirac spinors of
the baryons and the same constant $f_{\Lambda_c^+p}$, which encodes form-factor effects
at \,$\bar q^2=0$,\, enters both of the matrix elements, as explained in the Appendix.
It follows that
\begin{align}
{\cal M}_{\Lambda_c^+\to p\bar\gamma} & \,=\, 2 f_{\Lambda_c^+p}^{}\,
\overline{\textsl{\texttt U}_p^{}} \big( {\mathbb C} + \gamma_5^{} {\mathbb C}_5^{} \big)
i\sigma^{\mu\nu} \textsl{\texttt U}_{\Lambda_c}^{} \bar\varepsilon_\mu^*\bar q_\nu^{} \,, ~~~
\end{align}
which fulfills U(1)$_D$-gauge invariance, and so the branching fraction is
\begin{align} \label{BLc2pg'}
{\cal B}\big(\Lambda_c^+\to p\bar\gamma\big) & \,=\, \frac{\tau_{\Lambda_c^+}^{}\,
f_{\Lambda_c^+p}^2\, \big(m_{\Lambda_c^+}^2-m_p^2\big)\raisebox{1pt}{$^3$}}
{2\pi m_{\Lambda_c^+}^3} \big( |{\mathbb C}|^2 + |{\mathbb C}_5|^2 \big) \,, ~~~
\end{align}
where $\tau_{\Lambda_c^+}$ is the $\Lambda_c^+$ lifetime.
Numerically, we adopt \,$f_{\Lambda_c^+p}=0.50(1\pm0.07)$\, from the lattice QCD
calculation in Ref.\,\cite{Meinel:2017ggx}.
Incorporating its central value and those of $\tau_{\Lambda_c^+}$, $m_{\Lambda_c^+}$, and
$m_p^{}$ from Ref.\,\cite{Tanabashi:2018oca} into Eq.\,(\ref{BLc2pg'}), we then translate
the limit in Eq.\,(\ref{C2max}) into
\begin{align} &
{\cal B}\big(\Lambda_c^+\to p\bar\gamma\big) \,=\, 8.31\times10^{10}\, \tilde{\textsc c}{}^2
\,<\, 1.6\times10^{-5} \,. ~~~
\end{align}
If instead \,$f_{\Lambda_c^+p}=0.38(1\pm0.1)$,\, evaluated in the relativistic quark
model \cite{Faustov:2018dkn}, we would get a lower result,
\,${\cal B}\big(\Lambda_c^+\to p\bar\gamma\big)<9.1\times10^{-6}$.\,
These different numbers indicate the degree of uncertainty in the prediction.

In the case of \,$\Xi_c^0\to\Sigma^0\bar\gamma$,\, we employ the form-factor parameter
\,$f_{\Xi_c^0\Sigma^0}=0.46\pm0.12$,\, which has been estimated in the framework of light-cone
QCD sum rules \cite{Azizi:2011mw}.
Assuming isospin symmetry, we also have \,$f_{\Xi_c^+\Sigma^+}=f_{\Xi_c^0\Sigma^0}$\,
for \,$\Xi_c^+\to\Sigma^+\bar\gamma$.\,
Their branching fractions have expressions obtainable from Eq.\,(\ref{BLc2pg'}),
with suitable modifications.
With the central values of $f_{\Xi_c^0\Sigma^0}$ and the measured $\Xi_c^{+,0}$ lifetimes and
relevant baryon masses \cite{Tanabashi:2018oca}, along with Eq.\,(\ref{C2max}), we then arrive at
\begin{align}
{\cal B}\big(\Xi_c^+\to\Sigma^+\bar\gamma\big) & \,=\, 1.54\times10^{11}\, \tilde{\textsc c}{}^2
\,<\, 2.9\times10^{-5} \,, ~~~ \nonumber \\
{\cal B}\big(\Xi_c^0\to\Sigma^0\bar\gamma\big) & \,=\, 3.90\times10^{10}\, \tilde{\textsc c}{}^2
\,<\, 7.4\times10^{-6} \,. 
\end{align}
The difference between them is attributable mainly to \,$\tau_{\Xi_c^+}=3.9\,\tau_{\Xi_c^0}$.\,

\section{Decays of doubly charmed baryon\label{Xcc}}

In this and the next sections, we look at the impact of ${\cal L}_{uc\bar\gamma}$ on the decays of
hadrons containing two heavy quarks, specifically $\Xi_{cc}^{++}$ and $B_c^+$, respectively.
As LHCb will be the primary facility that produces them for the foreseeable
future~\cite{Cerri:2018ypt}, and they are either much less likely or too heavy to be produced
at ongoing $e^+e^-$ facilities, these transitions probably will not be probed any time soon.
We include them here for completeness, as the ingredients needed to treat them are already on hand.

Only one doubly charmed baryon, $\Xi_{cc}^{++}$, has been discovered so far \cite{Aaij:2017ueg},
with its lifetime and mass now also determined \cite{Tanabashi:2018oca}.
We take its spin parity, which is not yet established experimentally, to be \,$J^P=1/2^+$\,
based on quark-model expectations \cite{Aaij:2017ueg}.
Here we consider the decay channels
\,$\Xi_{cc}^{++}\to\Sigma_c(2455)^{++}\bar\gamma$\, and
\,$\Xi_{cc}^{++}\to\Sigma_c(2520)^{++}\bar\gamma$\, arising from ${\cal L}_{cu\bar\gamma}$,
as some information on the pertinent baryonic form factors has recently become
available~\cite{Hu:2020mxk}.

Since $\Sigma_c(2455)^{++}$ also has \,$J^P=1/2^+$ \cite{Tanabashi:2018oca}, the branching
fraction of \,$\Xi_{cc}^{++}\to\Sigma_c(2455)^{++}\bar\gamma$\, is analogous to that in
Eq.\,(\ref{BLc2pg'}), namely
\begin{align}
{\cal B}\big(\Xi_{cc}^{++}\to\Sigma_c(2455)^{++}\bar\gamma\big) & \,=\,
\frac{\tau_{\Xi_{cc}^{++}}^{}\, f_{\Xi_{cc}\Sigma_c(2455)}^2\, \big( m_{\Xi_{cc}}^2
- m_{\Sigma_c(2455)^{++}}^2 \big)\raisebox{1pt}{$^3$}}{2 \pi m_{\Xi_{cc}}^3}
\big( |{\mathbb C}|^2 + |{\mathbb C}_5|^2 \big) \,. ~~~
\end{align}
where $m_{\Xi_{cc}^{}}\equiv m_{\Xi_{cc}^{++}}$. 
In this equation, we use the form-factor parameter \mbox{$f_{\Xi_{cc}\Sigma_c(2455)}=-0.798$} 
which has been computed in the light-front quark model with an uncertainty of several
\mbox{percent \cite{Hu:2020mxk}}.
With the central values of the empirical $\Xi_{cc}^{++}$ lifetime and mass and
\mbox{$m_{\Sigma_c(2455)^{++}}=2453.97$\,MeV} from Ref.\,\cite{Tanabashi:2018oca},
plus the bound in Eq.\,(\ref{C2max}), we then find
\begin{align} &
{\cal B}\big(\Xi_{cc}^{++}\to\Sigma_c(2455)^{++}\bar\gamma\big) \,=\,
2.96\times10^{11}\, \tilde{\textsc c}{}^2 \,<\, 5.6\times10^{-5} \,. ~~~
\end{align}

In the second channel, the daughter baryon $\Sigma_c(2520)^{++}$ has spin parity \,$J^P=3/2^+$
\cite{Tanabashi:2018oca}.
As discussed in the Appendix, the baryonic matrix elements for this reaction are
\begin{align} \label{<Sc|usc|Xcc>}
\langle\Sigma_c(2520)^{++}|\overline ui\sigma^{\mu\nu} \big(1,\gamma_5^{}\big) c
|\Xi_{cc}^{++}\rangle\, \bar\varepsilon_\mu^* \bar q_\nu^{}
\,=\, \overline{\mbox{\small$\mathbb U$}}{}_{\Sigma_c}^\kappa \big(\gamma_5^{},1\big) &
\Bigg[ \big(\!\!\not\!\bar\varepsilon{}^*\bar q_\kappa^{}
- \bar\varepsilon_\kappa^{*\;} \slashed{\bar q} \big) {\cal F}
+ \frac{\bar q_\kappa^{}\!\!\not\!\bar\varepsilon{}^* \slashed{\bar q}\,
\tilde{\cal F}}{m_{\Xi_{cc}}} \Bigg] \textsl{\texttt U}_{\Xi_{cc}}^{} \,, ~
\end{align}
where $\mbox{\small$\mathbb U$}{}_{\Sigma_c}^\kappa$ denotes the Rarita-Schwinger spinor of
$\Sigma_c(2520)^{++}$ and the constants $\cal F$ and $\tilde{\cal F}$ parametrize form-factor
effects at \,$\bar q^2=0$.\,
We can then write the decay amplitude
\begin{align} \label{MXcc2Scg'}
{\cal M}_{\Xi_{cc}^{++}\to\Sigma_c(2520)^{++}\bar\gamma}^{} & \,=\, 2\,
\overline{\mbox{\small$\mathbb U$}}{}_{\Sigma_c}^\kappa \big( \gamma_5^{}{\mathbb C}
+ {\mathbb C}_5^{} \big) \Bigg[ \big( \!\!\not\!\bar\varepsilon{}^*\bar q_\kappa^{}
- \bar\varepsilon_\kappa^{*\;} \slashed{\bar q} \big) {\cal F}
+ \frac{\bar q_\kappa^{}\!\!\not\!\bar\varepsilon{}^* \slashed{\bar q}\,
\tilde{\cal F}}{m_{\Xi_{cc}}} \Bigg] \textsl{\texttt U}_{\Xi_{cc}}^{} \,, ~~~
\end{align}
which is U(1)$_D$-gauge invariant and leads to the branching fraction\footnote{It is worth
noting that Eqs.\,\,(\ref{MXcc2Scg'}) and (\ref{BXcc2Scg'}) can be shown to be consistent
in form with the amplitude and branching fraction, respectively, of
\,$\Lambda_b\to\Lambda(1520)\gamma$,\, which involves the ordinary photon,
in \cite{Hiller:2007ur}, after interchanging the parity-conserving and -violating terms in
the amplitude, as $\Sigma_c(2520)^{++}$ and $\Lambda(1520)$ are opposite in parity.}
\begin{align} \label{BXcc2Scg'}
{\cal B}\big(\Xi_{cc}^{++}\to\Sigma_c(2520)^{++}\bar\gamma\big) & \,=\,
\frac{\tau_{\Xi_{cc}^{++}}\, \Delta^8 {\cal F}^2}{12\pi m_{\Xi_{cc}}^3 m_{\Sigma_c}^2} \Bigg(
1 + \frac{4m_{\Sigma_c}^2}{\Delta^2} + \frac{2\tilde{\cal F}}{\cal F}
+ \frac{\Delta^2\tilde{\cal F}{}^2}{m_{\Xi_{cc}}^2 {\cal F}^2}  \Bigg)
\big( |{\mathbb C}|^2 + |{\mathbb C}_5|^2 \big) \,,
\end{align}
where \,$m_{\Sigma_c}\equiv m_{\Sigma_c(2520)^{++}}$\, and
\,$\Delta=\big(m_{\Xi_{cc}}^2-m_{\Sigma_c}^2\big)\raisebox{1pt}{$^{1/2}$}$.\,
Adopting \,${\cal F}=-0.635$\, and \,$\tilde{\cal F}=0.330$,\, which have been estimated in
the light-front quark model with uncertainties of several percent~\cite{Hu:2020mxk}, with
\,$m_{\Sigma_c}=2518.41$\,MeV \cite{Tanabashi:2018oca} and  Eq.\,(\ref{C2max}), we then obtain
\begin{align} &
{\cal B}\big(\Xi_{cc}^{++}\to\Sigma_c(2520)^{++}\bar\gamma\big) \,=\,
1.12\times10^{11}\, \tilde{\textsc c}{}^2 \,<\, 2.1\times10^{-5} \,. ~~~
\end{align}

\section{Decay of bottom, charmed meson\label{Bcmeson}}

The \,$c\to u\bar\gamma$\, transition can also bring about the decay of the bottom, charmed
pseudoscalar-meson $B_c^+$ into the bottom charged vector-meson $B^{*+}$ and the dark photon.
Consequently, the amplitude for \,$B_c^+\to B^{*+}\bar\gamma$\, and its branching fraction
are similar to the corresponding quantities associated with \,$D^+\to\rho^+\bar\gamma$,\,
given in Eqs.\,\,(\ref{MD2rg}) and (\ref{BD2rg}).
Especially, based on the latter we have
\begin{align} \label{BBc2B*g}
{\cal B}\big(B_c^+\to B^{*+}\bar\gamma\big) & \,=\, \frac{\tau_{B_c^+}^{}\, f_{B_c^{}B_u^*}^2
\big(m_{B_c^+}^2-m_{B^{*+}}^2\big)\raisebox{1pt}{$^3$}}{2 \pi m_{B_c^+}^3}
\big( |{\mathbb C}|^2 + |{\mathbb C}_5|^2 \big) \,. ~~~
\end{align}

To evaluate this, we employ the known $B_c^+$ lifetime and mass and
\,$m_{B^{*+}}=5324.70(22)$\,MeV from Ref.\,\cite{Tanabashi:2018oca}, as well as
\,$f_{B_c^{}B_u^*}^{} = 0.23\pm0.04$\, which has been calculated in Ref.\,\cite{Aliev:2006vs}
with QCD sum rules and is in agreement
with the older result \,$f_{B_c^{}B_u^*}=0.24$\, of Ref.\,\cite{Fajfer:1999dq} from
an application of the constituent quark model of Ref.\,\cite{Isgur:1988gb}.
Combining their central values and Eq.\,(\ref{C2max}) with Eq.\,(\ref{BBc2B*g}) then yields
\begin{align}
{\cal B}\big(B_c^+\to B^{*+}\bar\gamma\big) & \,=\, 3.54\times10^{10}\, \tilde{\textsc c}{}^2
\,<\, 6.7\times10^{-6} \,. ~~~~~
\end{align}

\section{Conclusions\label{concl}}

If a new U(1) gauge symmetry under which all SM particles are singlets exists and is
spontaneously broken, the accompanying dark photon has a nonzero mass and may undergo direct
renormalizable interactions with SM fermions due to kinetic mixing between the dark and SM
U(1) gauge fields.
This kind of dark photon has been the subject of many searches in recent years,
which have come up empty so far.
If the extra U(1) gauge group stays unbroken, the associated dark photon is massless instead and
can couple to SM members only through higher-dimensional operators, implying that it would
have evaded the aforesaid hunts for its massive counterpart.
Therefore, it is crucial that upcoming endeavors to seek dark photons take into account
the possibility that they are massless, in which case they may have consequential FCNC
interactions with SM fermions.

In this paper we explore the latter scenario, particularly that where the massless dark
photon has nonnegligible flavor-changing dipole-type couplings to the $u$ and $c$ quarks.
If occurring in nature, these interactions will induce the FCNC decays of charmed hadrons into
a~lighter hadron plus missing energy carried away by the dark photon.
We propose to pursue a number of such \,$\Delta C=1$\, reactions, focusing on those in which
the parent particles are the charmed pseudoscalar-mesons $D^{+,0}$ and $D_s^+$, the singly 
charmed baryons $\Lambda_c^+$, $\Xi_c^+$, and $\Xi_c^0$, the doubly-charmed 
baryon~$\Xi_{cc}^{++}$, and the bottom, charmed pseudoscalar-meson $B_c^+$.
We also \mbox{consider \,$D^0\to\gamma\bar\gamma$,\,} which has the standard photon besides 
$\bar\gamma$ in the final state.
In the context of a simplified new-physics model, we show that some of the $D^{+,0}$, $D_s^+$, 
and singly-charmed-baryon modes have branching fractions that are permitted by present 
constraints to be as large as several times $10^{-5}$, which may be within the sensitivity 
reach of BESIII and Belle II.
Since the same one or two $c\to u\bar\gamma$ operators are responsible for all these decays,
detecting one of them automatically implies specific predictions for the others,
allowing for additional empirical tests on the dark photon.
Conversely, measuring a bound on one of the decays will translate into expected bounds
on the others.
The results of this work will hopefully help motivate dedicated attempts to look for
massless dark photons via charmed-hadron processes.
These efforts will be complementary to the quests for the massive dark photon, which is
phenomenologically quite distinct from the massless one.

\acknowledgements

We would like to thank Cong Geng, Hai-Bo Li, and Min-Zu Wang for information on experimental
matters.
This research was supported in part by the MOST (Grant No. MOST 106-2112-M-002-003-MY3).

\appendix

\section{Baryonic matrix elements of \boldmath$\overline u\sigma^{\mu\nu}(1,\gamma_5)c$\label{bme}}

For the \,$\Delta C=1$\, transition between baryons \mbox{\small$\mathfrak B$} and
$\mbox{\small$\mathfrak B$}'$ with spin parity \,$J^P=1/2^+$,\, the baryonic matrix element of
\,$\overline u\sigma^{\mu\nu}c$\, has the general expression \cite{Chen:2001zc,Wang:2008sm} 
which is as follows:
\begin{align} \label{<usc>}
\langle\mbox{\small$\mathfrak B$}'(k)|\overline u\sigma^{\mu\nu}c
|\mbox{\small$\mathfrak B$}(k+\bar q)\rangle \,=\, \overline{\textsl{\texttt U}\,'} \Big(
\sigma^{\mu\nu} f_1^{} - \gamma^{[\mu} k^{\nu]}\, if_2^{} - \gamma^{[\mu}\bar q^{\nu]}\, if_3^{}
- k^{[\mu}\bar q^{\nu]}\, if_4^{} \Big) \textsl{\texttt U} \,, ~~~
\end{align}
where $k+\bar q$ and $k$ are the momenta of \mbox{\small$\mathfrak B$} and
$\mbox{\small$\mathfrak B$}'$, respectively, $\textsl{\texttt U}$ and $\textsl{\texttt U}\,'$
designate their Dirac spinors, $f_{1,2,3,4}^{}$\, denote form factors which are functions of
$\bar q^2$, and
\,${\tt X}^{[\mu}{\tt Y}^{\nu]}\equiv{\tt X}^\mu{\tt Y}^\nu-{\tt X}^\nu{\tt Y}^\mu$.\,
With the aid of the identity
\,$2i\sigma_{\eta\kappa}\gamma_5^{}=\epsilon_{\eta\kappa\nu\tau}\sigma^{\nu\tau}$\,
for \,$\epsilon_{0123}^{}=+1$,\, we can derive from Eq.\,(\ref{<usc>}) the corresponding matrix
element of \,$\overline u\sigma_{\eta\kappa}\gamma_5^{}c$,\,
\begin{align} \label{<usg5c>}
\langle\mbox{\small$\mathfrak B$}'(k)|\overline u\sigma_{\eta\kappa}^{}\gamma_5^{}c
|\mbox{\small$\mathfrak B$}(k+\bar q)\rangle
& \,=\, \overline{\textsl{\texttt U}\,'} \bigg[ \sigma_{\eta\kappa}^{}\gamma_5^{} f_1^{}
- \epsilon_{\eta\kappa\nu\tau}^{} \bigg( \gamma^{[\nu} k^{\tau]}\, \frac{f_2^{}}{2}
+ \gamma^\nu\bar q^\tau f_3^{} + k^\nu\bar q^\tau f_4^{} \bigg) \bigg] \textsl{\texttt U} \,.
\end{align}

In the treatment of the amplitude for
\,$\mbox{\small$\mathfrak B$}\to\mbox{\small$\mathfrak B$}'\bar\gamma$,\,
the matrix elements in Eqs.\,\,(\ref{<usc>}) and (\ref{<usg5c>}) are to be contracted with
the dark photon's polarization vector $\bar\varepsilon$ and momentum $\bar q$.
Thus, imposing \,$\bar\varepsilon\cdot\bar q=0$\, and
\,$\bar q^2=\slashed{\bar q}\slashed{\bar q}=0$,\, after straightforward algebra we arrive at
\begin{align} \label{<f|usc|i>}
\langle\mbox{\small$\mathfrak B$}'(k)|\overline u\sigma^{\mu\nu}c
|\mbox{\small$\mathfrak B$}(k+\bar q)\rangle\, \bar\varepsilon_\mu^* \bar q_\nu^{}
& \,=\, \overline{\textsl{\texttt U}\,'} \Bigg\{ \sigma^{\mu\nu} f_1^{(0)} - i \big[
\gamma^\mu \gamma^\nu \big(\slashed k+\slashed{\bar q}\big)
- \slashed k \gamma^\mu \gamma^\nu \big] \frac{f_2^{(0)}}{2} \Bigg\} \textsl{\texttt U}\,
\bar\varepsilon_\mu^*\bar q_\nu^{}
\nonumber \\ & \,=\, f_{\mathfrak{BB'}}^{}\, \overline{\textsl{\texttt U}\,'}
\sigma^{\mu\nu} \textsl{\texttt U}\, \bar\varepsilon_\mu^*\bar q_\nu^{} \,,
\\ \label{<f|usg5c|i>}
\langle\mbox{\small$\mathfrak B$}'(k)|\overline u\sigma^{\mu\nu}\gamma_5^{}c
|\mbox{\small$\mathfrak B$}(k+\bar q)\rangle\, \bar\varepsilon_\mu^* \bar q_\nu^{}
& \,=\, \overline{\textsl{\texttt U}\,'} \Bigg\{ \sigma^{\mu\nu}\gamma_5^{}\, f_1^{(0)}
- \epsilon^{\mu\nu\kappa\tau} \big[ \gamma_\kappa^{} \gamma_\tau^{}
\big(\slashed k+\slashed{\bar q}\big) - \slashed k \gamma_\kappa^{} \gamma_\tau^{} \big]
\frac{f_2^{(0)}}{4} \Bigg\}^{\vphantom{\int^{\int^|}}}
\textsl{\texttt U}\, \bar\varepsilon_\mu^* \bar q_\nu^{} ~~~
\nonumber \\ & \,=\, f_{\mathfrak{BB'}}^{}\, \overline{\textsl{\texttt U}\,'}
\sigma^{\mu\nu} \gamma_5^{} \textsl{\texttt U}\, \bar\varepsilon_\mu^* \bar q_\nu^{} \,,
\end{align}
where $f_{1,2}^{(0)}$ stand for $f_{1,2}^{}$ evaluated at \,$\bar q^2=0$\, and
\begin{align}
f_{\mathfrak{BB'}}^{} & \,=\, f_1^{(0)}
+ \frac{f_2^{(0)}}{2} (m_{\mathfrak B'}-m_{\mathfrak B}) \,, ~~~
\end{align}
with $m_{{\mathfrak B}{}^{(\prime)}}$ being the mass of $\mbox{\small$\mathfrak B$}{}^{(\prime)}$.
These results lead to Eq.\,(\ref{<p|usc|Lc>}).
Evidently, the same combination, $f_{\mathfrak{BB'}}^{}$, of form factors at \,$\bar q^2=0$\,
enters both Eqs.\,\,(\ref{<f|usc|i>}) and (\ref{<f|usg5c|i>}), in agreement with what was
previously found \cite{Hiller:2001zj,Gutsche:2013pp} concerning the baryonic matrix elements of
these types of tensor operators.

In the case where $\mbox{\small$\mathfrak B$}'$ is replaced by a spin-3/2 baryon
\mbox{\small$\mathbb B$}$'$, we can write
\begin{align}
\langle\mbox{\small$\mathbb B$}'(k)|\overline ui\sigma^{\mu\nu}c|{\mathfrak B}(k+\bar q)\rangle
\,=\; & \Big\{ \overline{\mbox{\small$\mathbb U$}}{}^{[\mu}\gamma^{\nu]} g_1^{}
+ \overline{\mbox{\small$\mathbb U$}}{}^{[\mu}k^{\nu]} g_2^{}
+ \overline{\mbox{\small$\mathbb U$}}{}^\eta\bar q_\eta^{} \Big( i\sigma^{\mu\nu} g_3^{}
+ \gamma^{[\mu}k^{\nu]} g_4^{} \Big) ~~~
\nonumber \\ & +\,
\overline{\mbox{\small$\mathbb U$}}{}^{[\mu}\bar q^{\nu]} g_5^{}
+ \overline{\mbox{\small$\mathbb U$}}{}^\eta\bar q_\eta^{} \Big( \gamma^{[\mu}\bar q^{\nu]} g_6^{}
+ k^{[\mu}\bar q^{\nu]} g_7^{} \Big) \Big\} \gamma_5^{} \textsl{\texttt U} \;,
\end{align}
where $k$ and $\mbox{\small$\mathbb U$}{}^\mu$ are, respectively, the momentum of
\mbox{\small$\mathbb B$}$'$ and its Rarita-Schwinger spinor, the latter satisfying the requirements
\,$\gamma_\mu\mbox{\small$\mathbb U$}{}^\mu=k_\mu\mbox{\small$\mathbb U$}{}^\mu=0$,\,
and $g_{1,2,\cdots,7}^{}$ represent form factors which depend on $\bar q^2$.
It follows that
\begin{align} \label{<3/2|usc|1/2>}
\langle\mbox{\small$\mathbb B$}'(k)|\overline ui\sigma^{\mu\nu}c|{\mathfrak B}(k+\bar q)\rangle\,
\bar\varepsilon_\mu^* \bar q_\nu^{}
\,= &~\, \overline{\mbox{\small$\mathbb U$}}{}^{[\mu}\gamma^{\nu]}\, \bar\varepsilon_\mu^*
\bar q_\nu^{}  \Bigg[ g_1^{(0)} + \frac{g_2^{(0)}}{2} \big(m_{\mathbb B'}^{}-m_{\mathfrak B}^{}\big)
\Bigg] \gamma_5^{} \textsl{\texttt U}
\nonumber \\ & +\,
\overline{\mbox{\small$\mathbb U$}}{}^\eta\bar q_\eta^{} \!\not\!\bar\varepsilon{}^*
\slashed{\bar q} \Bigg[ \frac{g_2^{(0)}}{2} - g_3^{(0)} - \frac{g_4^{(0)}}{2}
\big(m_{\mathbb B'}^{}+m_{\mathfrak B}^{}\big) \Bigg] \gamma_5^{} \textsl{\texttt U} ~~~~~
\nonumber \\ \,= &~\,
\overline{\mbox{\small$\mathbb U$}}{}^\eta \Bigg[ \big(
\bar\varepsilon_\eta^{*\;}\slashed{\bar q}\, - \!\not\!\bar\varepsilon{}^* \bar q_\eta^{}
\big) {\cal F} + \frac{\bar q_\eta^{}\!\!\not\!\bar\varepsilon{}^* \slashed{\bar q}\,
\tilde{\cal F}}{m_{\mathfrak B}^{}} \Bigg] \gamma_5^{} \textsl{\texttt U} \;,
\end{align}
where $g_j^{(0)}$ denotes the value of $g_j^{}$ at \,$\bar q^2=0$\, and
\begin{align}
{\cal F} & \,=\, g_1^{(0)} + \frac{g_2^{(0)}}{2}
\big(m_{\mathbb B'}^{}-m_{\mathfrak B}^{}\big) \,, &
\frac{\tilde{\cal F}}{m_{\mathfrak B}^{}} & \,=\, \frac{g_2^{(0)}}{2} - g_3^{(0)}
- \frac{g_4^{(0)}}{2} \big(m_{\mathbb B'}^{}+m_{\mathfrak B}^{}\big) \,, ~~~~~
\end{align}
with $m_{\mathbb B'}^{}$ being the mass of $\mbox{\small$\mathbb B$}'$.
Moreover,
\begin{align} \label{<3/2|usg5c|1/2>}
\langle\mbox{\small$\mathbb B$}'(k)|\overline ui\sigma^{\mu\nu}\gamma_5^{}c
|{\mathfrak B}(k+\bar q)\rangle\, \bar\varepsilon_\mu^* \bar q_\nu^{}
\,= &~\, \frac{\epsilon^{\mu\nu\kappa\tau}}{2}\, \langle\mbox{\small$\mathbb B$}'(k)|\overline u
\sigma_{\kappa\tau}^{}c|{\mathfrak B}(k+\bar q)\rangle\, \bar\varepsilon_\mu^* \bar q_\nu^{} ~~~~~
\nonumber \\ \,= &~\,
\overline{\mbox{\small$\mathbb U$}}{}^\eta \Bigg[ \big(\!\!\not\!\bar\varepsilon{}^*\bar q_\eta^{}
- \bar\varepsilon_\eta^{*\;} \slashed{\bar q} \big) {\cal F}
+ \frac{\bar q_\eta^{}\!\!\not\!\bar\varepsilon{}^* \slashed{\bar q}\,
\tilde{\cal F}}{m_{\mathfrak B}^{}} \Bigg] \textsl{\texttt U} \,.  
\end{align}
From Eqs.\,\,(\ref{<3/2|usc|1/2>}) and (\ref{<3/2|usg5c|1/2>}), we get
Eq.\,(\ref{<Sc|usc|Xcc>}).

In the derivation of the \,$\mbox{\small$\mathfrak B$}\to\mbox{\small$\mathbb B$}'\bar\gamma$\, rate,
the absolute square of the decay amplitude needs to be summed over the initial and final
particles' polarizations.
The expression for the sum over the \mbox{\small$\mathbb B$}$'$ polarizations is available from
the literature ({\sl e.g.}\,\cite{Christensen:2013aua}).
It is given here for completeness:
\begin{align} \label{pols}
\raisebox{2pt}{\footnotesize$\displaystyle\sum_{\varsigma=-3/2}^{3/2}$}\,
\mbox{\small$\mathbb U$}{}_\varsigma^\beta(k)\,
\overline{\mbox{\small$\mathbb U$}}{}_\varsigma^\eta(k)
& \;=\; \big(\slashed k + m_{\mathbb B'}^{}\big) \bigg( \frac{\gamma_\mu^{}\gamma_\nu^{}}{3}\,
{\cal G}^{\mu\beta}(k)\, {\cal G}^{\nu\eta}(k) - {\cal G}^{\beta\eta}(k) \bigg) \,,
\end{align}
where \,${\cal G}^{\mu\nu}(k)=g^{\mu\nu}-k^\mu k^\nu/m_{\mathbb B'}^2$.\,

\end{document}